\documentstyle[sprocl]{article}

\input{psfig.sty}

\def\be{\begin{equation}}
\def\ee{\end{equation}}
\def\ba{\begin{array}}
\def\ea{\end{array}}
\def\bea{\begin{eqnarray}}
\def\eea{\end{eqnarray}}

\begin{document}

\title{Soft Color Fields in DIS at low $x$ and low $Q^2$}

\author{A. Metz}

\address{CEA-Saclay, DAPNIA/SPhN, F-91191 Gif-sur-Yvette, France
\\E-mail: ametz@cea.fr}

\maketitle

\abstracts{Two complementary approaches to DIS at low $x$ and low $Q^2$
are presented.
In the first case we apply a model containig two Pomeron trajectories.
In the second case we determine the gluon density in the semiclassical
treatment at next-to-leading order.
Both approaches rely on the concept of soft color fields.}

\section{Introduction}
Regge-theory turned out to be very sucessful in describing the rise of
hadronic cross sections by means of a so-called soft Pomeron trajectory.
The same picture works well also for real photoabsorption on the proton.
However, in the case of DIS with an incoming virtual photon and 
virtualities $Q^2 \geq 1\,\textrm{GeV}^2$, HERA data show a clear 
deviation from a soft Pomeron behaviour of the cross section.
Therefore, Donnachie and Landshoff \cite{donnachie_98} recently proposed 
a two-component model containig a soft and a hard Pomeron trajectory, 
where the influence of both components varies with $Q^2$.
In the first approach \cite{dalesio_99} discussed below we also exploit 
a model with two Pomerons relating the soft Pomeron to non-perturbative 
QCD by means of the Model of the Stochastic Vacuum (MSV) of Dosch and 
Simonov \cite{dosch_87}. 
This model can be considered as approximation to QCD in the infrared 
regime and provides, together with an ansatz for the quark wave 
function of the proton, a specific description of the soft color field
of the proton.
Our hard Pomeron is related to perturbative QCD.

In the second approach \cite{dosch_99} we focus on the concept of parton 
densities frequently used in the analysis of DIS. 
More precisely, we determine in the semiclassical 
approach \cite{buchmueller_97,buchmueller_99} the gluon density 
at next-to-leading order (NLO), which can serve as input in the evolution 
equations.
In the semiclassical treatment of DIS the interaction of the partonic 
fluctuation of the virtual photon with the soft color field, describing the
target at low $x$, is treated in an eikonal 
approximation \cite{nachtmann_91}. 
Our result is quite general.
The gluon density is expressed in terms of a (non-perturbative) Wilson
loop and can be evaluated in any model of the soft color field of the 
target. 
On the other side, the concept of parton densities in general can not be 
continued to the real photon point.
In that sense our second approach is more limited than the Pomeron model.

\section{Soft and Hard Pomeron in DIS}

For the structure function $F_2$ a model based on two Pomerons 
reads \cite{donnachie_98} 
\begin{equation} \label{f2_b}
F_2(x,Q^2) = f_s(Q^2) x^{-\lambda_s} + f_h(Q^2) x^{-\lambda_h},
\end{equation} 
where the exponents $\lambda_s$ and $\lambda_h$ of the soft and the hard 
contribution respectively do not depend on $Q^2$.
We use $\lambda_s = 0$, and $\lambda_h$ is treated as a free parameter.
Though we can not explain the Pomeron intercepts, the residue functions 
$f_s$ and $f_h$ are related to QCD.
\\

We evaluate $f_s$ in the MSV, which is a specific model of 
non-perturbative QCD derived from the assumption that the infrared 
behaviour of QCD can be approximated by a Gaussian stochastic 
process \cite{dosch_87}.
In terms of the Wilson area law the MSV predicts linear confinement.
The correlator of the gluon field strength, which has been computed on 
the lattice \cite{digiacomo_92}, serves as central quantity of the MSV. 
The lattice simulation of the correlator shows a transition between 
non-perturbative and perturbative effects roughly at the correlation
length $a \approx 0.3\,\textrm{fm}$.
In the following we will exploit the scale $a$ to separate soft and hard 
contributions.
\\
In the eikonal approximation the cross section $\sigma_{q\bar{q}}$ for 
scattering a color dipole off the proton is energy independent in the 
MSV \cite{dosch_94}.
With the wave function $\Psi_{L,T}(Q^2,z,r)$ describing the fluctuation
of a longitudinal or transverse photon into a $q\bar{q}$ pair, the soft 
part of the photoabsorption cross section on the proton takes the form
\begin{equation} \label{dipol}
\sigma_{L,T}^{soft}(Q^2) = 2\pi \int_0^1 dz \int_a^{\infty} dr\,r 
 |\Psi_{L,T}(Q^2,z,r)|^2 \sigma_{q\bar{q}}^{MSV}(z,r) \,,
\end{equation}
where $z$ is the longitudinal momentum fraction of the quark and $r$ the 
transverse size of the fluctuation.
To ensure confinement we use an effective quark mass \cite{dosch_98} 
interpolating between a constituent quark at low $Q^2$ and a current 
quark at high $Q^2$.
In Eq.~(\ref{dipol}) we introduce the lower cutoff $a$ in the $r$ 
integration \cite{rueter_99}, since the interaction at small distances 
is given by the hard component.
According to $F_2^{soft} = Q^2(\sigma_L^{soft} + \sigma_T^{soft})
/4\pi^2\alpha_{QED}$, we can compute the soft contribution 
from (\ref{dipol}).
The only free parameter of $F_2^{soft}$ is the constituent quark 
mass \cite{dalesio_99}.
\begin{figure}[tbp]
\vspace{-0.1in}
\hspace{2.5cm}
\centerline{\hbox{
\psfig{file=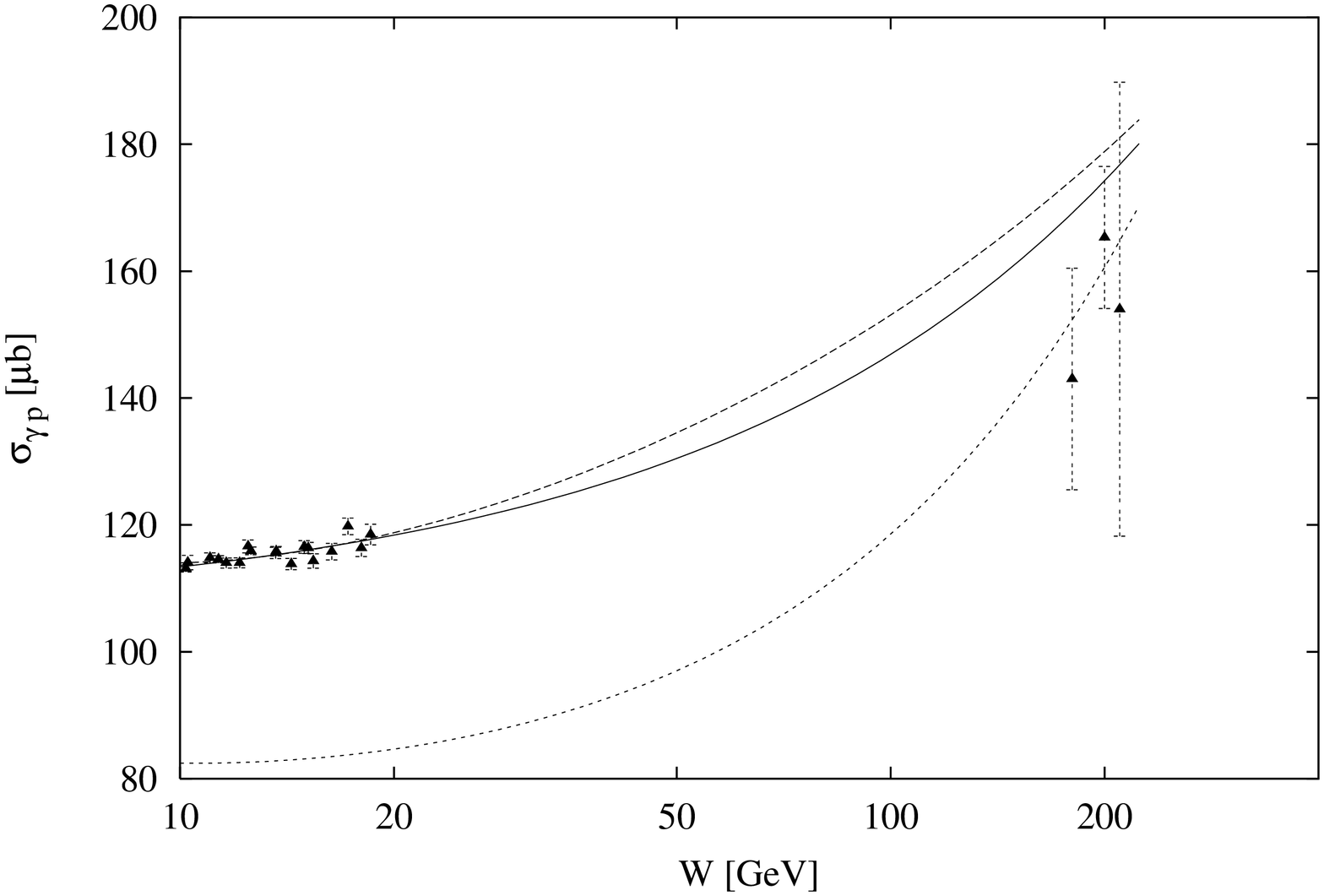,height=1.8in,width=2.45in,angle=-90}
\psfig{file=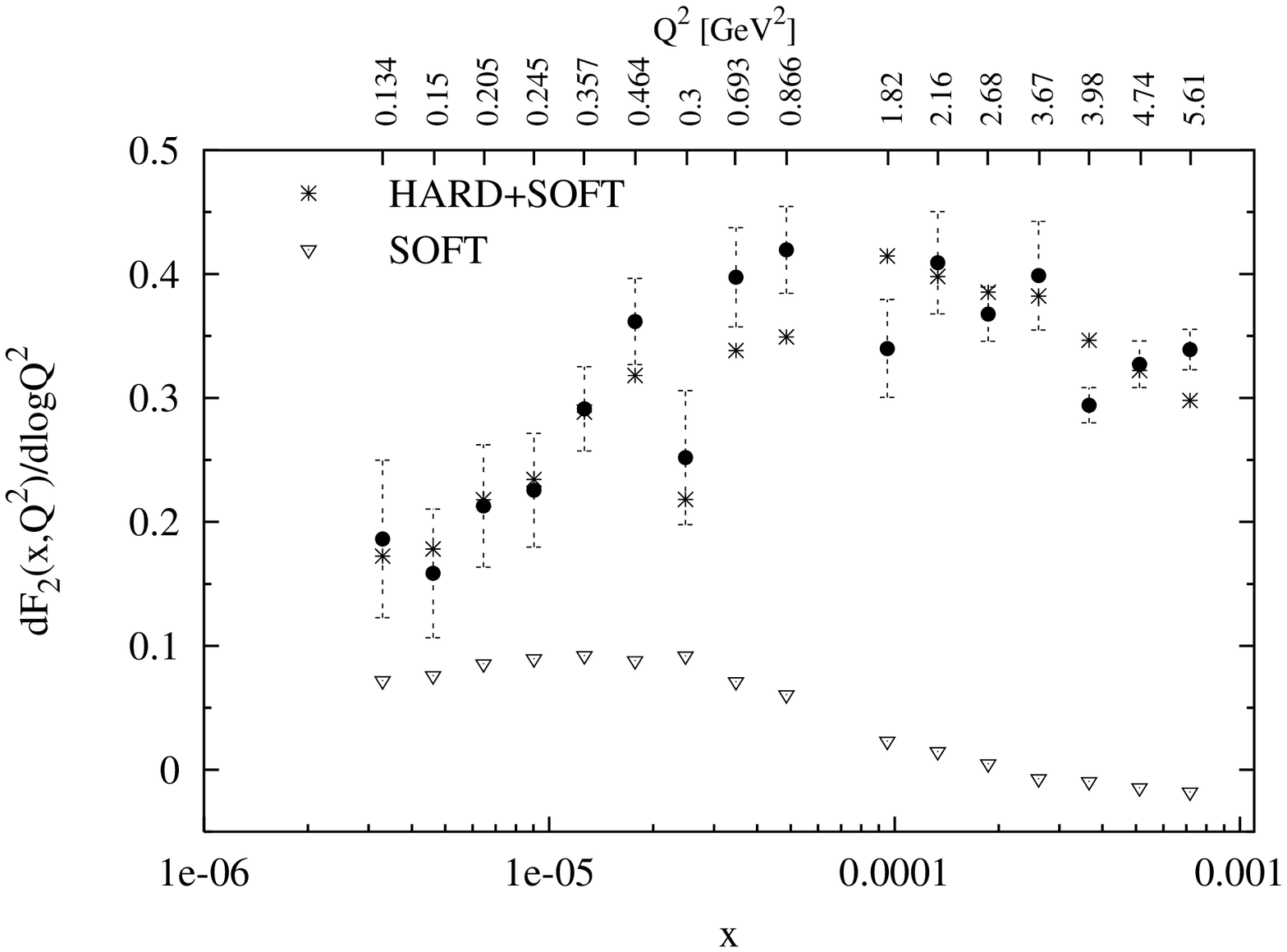,height=2.0in,width=2.45in,angle=-90}
}}
\parbox{12cm}
\caption{\label{res} 
Left panel: Total cross section for real photoproduction. 
Our fit (full line) is compared to those performed in 
Ref. \cite{donnachie_98} (dashed line) and in Ref. \cite{adel_97} 
(dotted line).
Right panel: Logarithmic derivative of $F_2$ vs $x$.} 
\end{figure}
\\

To model $F_2^{hard}$ we consider as starting point the evolution of 
a power-behaved $F_2$ derived by L\'opez and Yndur\'ain \cite{lopez_80}.
To leading order perturbative QCD implies that the singlet structure 
function reads
\begin{equation} \label{f2_pert}
F_2^{pert}(x,Q^2) = C_2 \, \alpha_s(Q^2)^{-d_{+}(1+\lambda)} 
x^{-\lambda} \,,
\end{equation}
with $d_{+}$ denoting the leading eigenvalue of the anomalous dimension
matrix of the quark-singlet and gluon evolution kernel.
The quantities $C_2$ and $\lambda$ are free parameters.
Eq.~(\ref{f2_pert}) is based on a singular gluon input and only valid at 
low $x$.
\\
To obtain a hard component, which is suitable also at low 
values of $Q^2$, we multiply $F_2^{pert}$ in (\ref{f2_pert}) by a 
phenomenological factor and freeze the strong coupling. 
In doing so we introduce a further parameter ($M$), and have
\begin{eqnarray}
F_2^{hard}(x,Q^2) & = &
 C_2 \, \tilde{\alpha}_s(Q^2)^{-d_{+}(1+\lambda)}
 x^{-\lambda} \bigg( \frac{Q^2}{Q^2+M^2} \bigg)^{1+\lambda} ,
 \\
& & \textrm{with}\quad \tilde{\alpha}_s(Q^2) 
 = \frac{4\pi}{\beta_0 \,\ln ((Q^2+M^2)/\Lambda_{QCD}^2)} \,.
\nonumber
\end{eqnarray}
In particular, $F_2^{hard}$ leads to a finite cross section 
$\sigma_{\gamma p}$ for real photoproduction.
A different modification of $F_2^{pert}$ in the region of low $Q^2$ has 
been proposed in Ref. \cite{adel_97}.
Our complete ansatz for $F_2$ is given by the sum of $F_2^{soft}$ and
$F_2^{hard}$. 
\\

The four parameters of the model are fitted to data for real \cite{real}
and virtual \cite{virtual} photoabsorption, where the kinematical cuts 
$Q^2 \leq 6.5\,\textrm{GeV}^2$, $x \leq 0.01$ and 
$W \geq 10\,\textrm{GeV}$ have been used to select the data.
We obtain a $\chi^2/\textrm{d.o.f} = 0.98$ for 222 data points and
the exponent $\lambda_h = 0.37$.
\\
On the l.h. side of Fig.~1 we have plotted our $\sigma_{\gamma p}$ in 
comparison with the fit of Donnachie and Landshoff \cite{donnachie_98}, 
and Adel, Barreiro and Yndur\'ain \cite{adel_97}.
The parametrization of Ref.~\cite{adel_97} has similarities to our 
approach but significantly underestimates the low energy data, which 
were not included in the fit.
The r.h. side of Fig.~2 shows the logarithmic derivative of $F_2$
(Caldwell plot \cite{caldwell_97}).
This picture demonstrates that in particular the turnover in the Caldwell 
plot can be described by a two-Pomeron model.
\\
The extension of our model to higher values of $Q^2$ still has to be 
analysed.
In addition, one has to investigate the consequences if the soft 
contribution is multiplied by the energy dependence of the soft Pomeron
of hadron scattering.
\section{The Semiclassical Gluon Distribution at Next-to-Leading Order}
In the semiclassical approach, one considers, at low $x$, the proton 
as localized soft color field without specifying this field. 
DIS is treated in the target rest frame, where the photon acquires a
partonic fluctuation.
This picture of DIS allows for a combined description of both
inclusive and diffractive events \cite{hebecker_99}.
\\

For extracting the gluon density it is convenient to use a `scalar 
photon' (denoted by $\chi$) coupled directly to the gluon 
field \cite{mueller_90} via the lagrangian 
\begin{equation}
{\cal L}_{I} = -\frac{\lambda}{2}\,\chi\,\mbox{tr} F_{\mu\nu}F^{\mu\nu} \,.
\end{equation}
The gluon density is derived by matching the semiclassical and the 
parton model approach.
This means that to leading order we have to equate the cross section for 
the transition $\chi \to g$ in an external field with the cross section 
of the process $\chi g \to g$ as given in the parton model, where the 
former is evaluated in the eikonal approximation.
To leading order one finds the result \cite{buchmueller_99}
\begin{eqnarray}
xg^{(0)}(x,\mu^2) & = & \frac{1}{12\pi^2\alpha_s} 
\int d^2x_{\perp} \left| \frac{\partial}{\partial y_\perp}
W^{\cal A}_{x_\perp}(y_\perp) \Big|_{y_\perp=0} \right|^2 \,,
\quad \textrm{where}
\\
W^{\cal A}_{x_\perp}(y_\perp) & = & 
U^{\cal A}(x_\perp) U^{{\cal A}\dagger}(x_\perp + y_\perp) - 1
\vphantom{\frac{1}{1}} \nonumber 
\end{eqnarray}
is a Wilson loop in the adjoint representation, and the phase factor
\begin{equation}
U^{\cal A}(x_\perp) = P \exp \bigg[-\frac{ig}{2} 
\int_{-\infty}^{\infty} dx_{+} \, A_{-}^{\cal A}(x_{+},x_{\perp}) \bigg]
\end{equation}
governs the eikonalised interaction of a fast gluon in an external color
field.
The gluon distribution $xg^{(0)}(x,\mu^2)$ is a constant measuring the
averaged local field strength of the target.
Based on this leading order result, together with a logarithmic energy
dependence introduced by hand, a sucessful description of DIS data has 
been obtained \cite{buchmueller_99}.
\begin{figure}[htbp]
\centerline{\hbox{\psfig{file=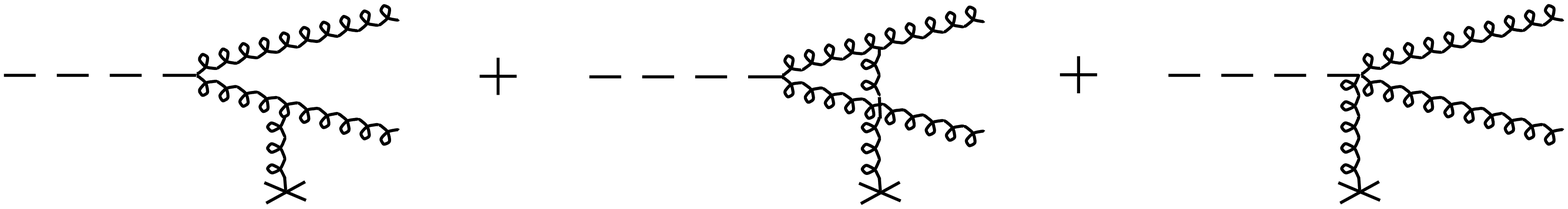,width=7.5cm}}}
\parbox{12cm}
\caption{\label{diag} 
Amplitude for $\chi \to gg$, which can be used to calculate the total 
semiclassical cross section in the high energy limit at NLO.}
\end{figure}

At NLO, we write the gluon density as
\begin{equation} \label{gd_sum}
xg(x,\mu^2) = xg^{(0)}(x,\mu^2) + xg^{(1)}(x,\mu^2)\,,
\end{equation}
with $xg^{(1)}(x,\mu^2)$ denoting the (scheme dependent) NLO 
correction.
To extract this correction, the cross section for the transition 
$\chi \to gg$ in an external field has to be equated with the parton model
cross section of the process $\chi g \to gg$.
In the high energy limit, the total cross section $\chi \to gg$ 
\cite{dosch_99} can be obtained from the eikonalized version of the 
diagrams in Fig.~2.
Note that Fig.~2 just shows the leading contributions, which arise when
expanding the eikonalized amplitude in powers of the external field.
Because of the limited space, we quote here only the final result
of the gluon density at NLO \cite{dosch_99}, given in the 
$\overline{\mbox{MS}}$ scheme, without presenting any detail of the 
calculation:
\begin{eqnarray} \label{gd_nlo}
xg^{(1)}(x,\mu^2) & = & \frac{1}{\pi^3} \left( \ln\frac{1}{x} \right) \,
\int_{r(\mu)^2}^{\infty} \, \frac{dy_\perp^2}{y_\perp^4}
\left\{ -\int d^2x{_\perp} \,\mbox{tr} \, 
W^{\cal A}_{x_\perp}(y_{\perp}) \right\}\,,
\\
& & \mbox{with} \quad r(\mu)^2=
\frac{4e^{\frac{1}{12}-2\gamma_E}}{\mu^2}\,.
\nonumber
\end{eqnarray}
The scheme dependence of the gluon density enters through the 
short-distance cutoff $r(\mu)$.
At NLO, the gluon density shows a $\ln (1/x)$ enhancement at small $x$, 
and is sensitive to the large-distance structure of the target.
\\
If one exploits the model of a large hadron \cite{mclerran_94} to describe 
the color field of the proton, a comparison of our result with the one 
of Mueller \cite{mueller_90,mueller_99} becomes possible.
We find agreement for both the integrated distribution in (\ref{gd_nlo}) and 
the unintegrated gluon density \cite{dosch_99} not shown here.
However, in Refs. \cite{mueller_90,mueller_99}, where the main focus is
on parton saturation, the scale dependence of the gluon density has not 
been discussed.
More precisely, we provide for the first time a quantitative relation
between the short-distance cutoff in Eq. (\ref{gd_nlo}) 
and the scale of the gluon distribution, which can only be achieved by 
matching the semiclassical approach with a treatment in the parton model.
\\ 
The result (\ref{gd_nlo}) enables us to obtain numerical predictions for 
the gluon density at NLO in any non-perturbative approach describing 
the soft color field of the proton.
In future work a comparison with DIS data has to be done using the 
semiclassical NLO distribution as input for the NLO evolution 
equations.

\section*{Acknowledgments}
The results presented here emerged from an interesting and fruitful
collaboration with U. D'Alesio, H.G. Dosch, A. Hebecker and H.J. Pirner.
I am grateful to A. Hebecker for reading this manuscript.
This work has been supported by the European TMR programme 
''Hadronic Physics with High Energy Electromagnetic Probes''.

\end{document}